# Automated Assessment of Transthoracic Echocardiogram Image Quality Using Deep Neural Networks

Robert B. Labs[1], Apostolos Vrettos[2], Jonathan Loo[1], Massoud Zolgharni[1]
[1] School of Computing and Engineering, University of West London, London, UK
[2] Imperial College, Healthcare, NHS Trust, UK
robbie.labs@uwl.ac.uk

## ABSTRACT

**Background:** Standard views in two-dimensional echocardiography are well established but the quality of acquired images are highly dependent on operator skills and are assessed subjectively. This study is aimed at providing an objective assessment pipeline for echocardiogram image quality by defining a new set of domain-specific quality indicators. Consequently, image quality assessment can thus be automated to enhance clinical measurements, interpretation, and real-time optimization.

**Methods**: We have developed deep neural networks for the automated assessment of echocardiographic frame which were randomly sampled from 11,262 adult patients. The private echocardiography dataset consists of 33,784 frames, previously acquired between 2010 and 2020. Unlike non-medical images where full-reference metrics can be applied for image quality, echocardiogram's data is highly heterogeneous and requires blind-reference (IQA) metrics. Therefore, deep learning approaches were used to extract the spatiotemporal features and the image's quality indicators were evaluated against the mean absolute error. Our quality indicators encapsulate both anatomical and pathological elements to provide multivariate assessment scores for anatomical visibility, clarity, depth-gain and foreshortedness, respectively.

**Results**: The model performance accuracy yielded 94.4%, 96.8%, 96.2%, 97.4% for anatomical visibility, clarity, depth-gain and foreshortedness, respectively. The mean model error of 0.375±0.0052 with computational speed of 2.52 ms per frame (real-time performance) was achieved.

**Conclusion**: The novel approach offers new insight to objective assessment of transthoracic echocardiogram image quality and clinical quantification in A4C and PLAX views. Also lays stronger foundations for operator's guidance system which can leverage the learning curve for the acquisition of optimum quality images during transthoracic exam.

## 1. INTRODUCTION

A two-dimension (2D) echocardiogram has become de-facto of assessing cardiac functions because it presents rich anatomical details of the myocardium, and for its non-ionizing in-vivo advantages. Nevertheless, echocardiogram quality assessment is not void of technical and operational drawbacks. Firstly, echocardiogram images are produced through scattering centers and do not present crisp edges unlike the non-medical images. Secondly, the acquisition of high-quality echo images requires a significant experts' skill, and the standards of image quality are commonly exacerbated by user's subjective assessment and patients' anatomical profiles. For example, there is a strong indication why quantification of systolic function is recommended for apical-four (A4C) and parasternal long axis (PLAX) views [1], [2], [3], According to Nosir (1997) and Lang (2015), the spatial orientations of A4C and PLAX views are congruent in orientation, thus offering complementary advantages on the heart's functional measurement and quantifications. However, echocardiogram quality assessment still exists in subjective domain, with significant impacts, as documented in [4], [5], [6], ultimately pointed to the issues of clinical reliability, misdiagnosis, and poor response to patient care. These drawbacks remain significant and are inhibiting the adoption of echocardiograms as the reliable imaging modality despites its many advantages. Apparently, a good quality image provides a more accurate delineation of myocardial borders and yields accurate measurement [3], but the factor perceived as 'good quality' indicator is based on individual assessment which varies throughout the clinical practices. This indicates the need for a coherent standard and benchmark for determining the constituent of a 'good quality' image.

Currently, the method of echocardiographic image assessment entails manual inspection of echo images (sometimes large number of images) to determine its clinical and pathological relevance. This process is known to be time consuming, laboriously expensive, and precipitates variability of opinions on diagnostic outcomes. Consequently, an automated assessment is thus required for consistency, reliability, assessment, and optimization.

### 1.1 The Objective of this Study

The assessment of image quality in natural images (non-medical images) are quite straight forward with deep learning approaches, as they are modelled either with full-reference (FR-IQA) or reduce-reference (RR-IQA) metrics. On the other hand, 2D echocardiographic images (Fig. 2) are formed by interference pattern of scattering centres with inherent poor lateral and axial resolutions. Basically, the echocardiogram's anatomical features do not present crisp edges and boundary either because the endocardium is trabeculated with papillary muscles and the external purkinje networks [7]. The complexity of heart structure and the relationship between the epicardium, myocardium, and pericardium does exacerbate the acoustic impedance transition between its soft layers and precipitates the heterogeneity in echocardiograms. Due to this problem, it's grossly inadequate to model 2D echocardiogram under distortion-specific metric [8]. Echocardiogram present significantly subtle differences in its successive frames, which human eyes cannot detect [9]. This is a direct contrast to natural images which usually feature known distortion types.

The objective of this study, therefore, focuses on 2D echocardiographic image quality indicators and the method of its assessment. Although several research efforts have indicated the value of objective standard and proposed a number of assessment methods on how echocardiographic image quality [10], [11], [12] can be evaluated. Unfortunately, the efforts do not meet translatory acceptance, either because they are limited in scope, or they do not represent expert's clinical standard in objective benchmarking. Therefore, the implementation of an automated assessment protocol would be based on a fundamental definition of domain-specific objective standards and method of its assessment.

However, in the context of ultrasound, a coherent standard of objectivity relating to general echocardiography's digital image is essential but hard to define outside the clinical practice [6]. But its commonly admitted that the constituents of echocardiogram good-quality image should be relative or congruent to clinical significance and anatomical delineation. This knowledge comes with many years of experts' professional experiences. Hence, an objective standard should include specific elements, that purposely encapsulate both the anatomical and pathological visibility, with the possibility for quality optimization and real-time assessment. Consequently, by defining the element of image quality attributes (indicators) in these two planes, we can effectively model an inclusive, objective standards benchmark which are relevant to generalized clinical protocols in echocardiography. We believe this would provide objective arbitration to improve reliability of cardiac measurement, quantification, and diagnostic accuracy in echocardiography.

*1.2 Related Work*

All previously reported studies on objective assessment have used a limited criteria to define objective quality and deemed inappropriate of clinical assessment in clinical practice. Also, the method of assessment using a weighted average score index has been considered unsuitable for real-time acquisition and optimization guidance. Hence, the practical deployability of such a system is limited to experimental demonstration instead of translatory advantage. A clinically relevant system therefore would provide insight on objective standard and method of assessing specific quality attributes.

We have earlier demonstrated the feasibility of such a system of assessment [7], [13] and hereby provide details of wider implementations and how it can be clinically deployed in a unified workflow. Apart from validating an objective quality system with external clinical dataset, the proof of concept for operators' guidance on automated echocardiography have been discussed in many studies [14], [15], but implementing a useful pipeline for clinical advantages has achieved little impact. This is because a pipeline that offers weighted average scores is incapable of specificity in relation to the elements of image quality and operators' acquisition skills. All existing pipeline (without exception) that are capable of real-time feedback have only indicated the maximum quality ratings of echo images without suggesting 'how' or 'what' aspect of image requires an improvement. This means that operators are left to utilize their acquisition experience, which is of a little benefit to less experienced operators.

One of the earliest works on objective assessment of cardiac image quality is Abdi et al (2017). He demonstrated the feasibility of objective assessment using convolutional neural network models in five apical views using six (6) criteria scoring methods [16], [17]. Since there was no publicly available cardiac dataset to model, the author relied on expert's knowledge for its feature engineering, a high resource intensive process. Abdi's 85% of model accuracy was regarded as plausible outcome but was clinically deficient for transthoracic standard examination practice. This is because the defined quality indicators are limited and do not represent experts' global characteristics for cardiac diagnosis using 2D echocardiographic images.

Alternatively, Luong et al., (2020) research utilised twelve (12) criteria to grade each of the nine apical standard views, (Table III) while computing a continuous single variable score to represent objective quantity for respective apical views [18]. Luong's regression model achieved overall accuracy of 87% with regards to four expert ground truths and sufficiently demonstrated the impact of image quality on diagnostic utility. However, the assessment method and scores do not represent cardiologists' conventional assessment in practice, hence, cannot be applied in clinical workflow.

The most recent study on objective quality assessment by Dong et al., (2020) [14]. However, the study was limited to fetal ultrasound in apical four-chamber plane (A4C) and did not include PLAX view nor similar score criteria which can be independently accessed for adults' echocardiography examination in clinical practice. Dong's argument for focus/zoom attributes emanated from fetal cardiology where specific tissue became the focus of an investigation. However, these image attributes, though important, should be described as elements of clarity. Therefore, a zoomed section of myocardium should exhibit the attributes of clarity, instead of being considered as an independent indicator.

*1.3 Main Contributions*

In the light of the above related work, we admit the research efforts are plausible contributions, however, the specified criteria used to define quality assessment are limited in scope and are insufficient for TTE's clinical relevance. Also, the existing assessment methods do not match expert's expectations as currently obtainable in clinical practice.

However, in this research, we examined all existing quality criteria and proposed additional criteria that could translate to experts' subjective assessment. Finally, we defined, for the first time, a novel, most comprehensive criteria, and objective attributes (quality indicators) by which cardiac images can be assessed and optimized. We summarize our main contributions as follows:

- Demonstrate the feasibility of a novel, coherent and clinically relevant objective standard for the assessment of 2D echocardiographic images which account for relevant anatomical profiles, linear and volumetric quantifications of myocardial functions.
- Fresh insight to real-time assessment method that provides access to specific elements of cardiac image quality for the purpose of image optimization, accurate quantification, and diagnosis.
- Annotation of an independent echocardiography patient dataset showing four attributes of image quality namely: anatomical visibility, chamber clarity, depth-gain, and fore-shortening attributes for A4C, PLAX apical standard views.
- Public release of experts' annotated patient dataset to allow future studies and external validation of the new approaches or methods available on request at IntSav-QLabs [31]
- Detailed implementation of multi-stream deep learning architecture pipeline to process and allow access to specific image attributes in A4C and PLAX view of echo cine loop.

**2. MATERIALS AND METHODS**

We provide a detailed account of the dataset description, and justification for the collective elements required for objective standard assessment of image quality on A4C and PLAX views, and how they can be optimized in real-time deployment. This is followed by the expert annotation process and details for the implementation of our deep convolutional neural network model.

*2.1 Dataset Source and Ethical Approval*

At present, no echocardiogram dataset with the corresponding four separate annotations on A4C/PLAX image quality assessment is publicly available. We, therefore, aimed at preparing our own dataset (echocardiograms and corresponding ground-truth) for model developments. A large random sample of echocardiographic studies from different patients performed between 2010 and 2020 was extracted from Imperial College Healthcare NHS Trust's echocardiogram database. Ethical approval was obtained from the Health Regulatory Agency for the anonymized export of large quantities of imaging data. It was not necessary to approach patients individually for consent of data originally acquired for clinical purposes.

The images were acquired during examinations performed by experienced echocardiographers, according to the standard protocols for using ultrasound equipment from GE Healthcare and Philips Healthcare manufacturers. Automated anonymization was performed to remove the patient-identifiable information from DICOM-formatted videos.

A neural network model, previously developed in our research group [19], was then used to detect different echocardiographic views and separate the A4C and PLAX views. This resulted in a total of 33,784 frames from different patients: 15,476 and 18,308 frames for A4C and PLAX, respectively.

## *2.2 Definition and Grouping of Quality Attributes Indicators*

View-specific image quality scoring indicators (attributes) and criteria was defined by consulting our clinical expert committee at National Heart and Lung Institute. Four main quality attributes were considered for each view, which are listed in Table I, and enumerated as follows.

*2.2.1 Anatomical Visibility*

Unlike photographic images, ultrasound images are formed by interference patterns of scattering centers that do not present clear edges, but inherently poor lateral and axial resolutions [7]. Hence, the magnitude of visibility on chamber cavities for both A4C and PLAX frames can be correctly visualized using the correct method of heart's apex slicing, to yield the acceptable clinical projection of images' anatomical structures. This could present a sharp or blurred edges [15] of amplitude structures. Equations (1 - 2) describe the rotation of a frame vector in two-dimensional spatial distribution where $x_1 y_1$ represent on-axis projection, taking arbitrary center $x_c$, $y_c$, off-axis $x_p, y_p$ can thus be mitigated from $\beta$ known angle to improve anatomical visibility. In A4C, emphasis is placed on apical orientation, echogenicity of the left ventricle chamber, mitral and atrium valves [2]. Although the LV apex is not visualized in PLAX, but emphasis is placed on the anatomical echogenicity and clinical orientation of the right ventricle, left ventricle, the pericardium positions, and the aortic valves. These are clinically relevant features experts rely on for quantification, clinical assessment, and diagnosis.

$$x_1 = (x_p - x_c)\cos\beta - (y_p - y_c)\sin\beta + x_c \quad (1)$$

$$y_1 = (x_p - x_c)\sin\beta - (y_p - y_c)\cos\beta + y_c \quad (2)$$

*2.2.2 Cavity Clarity*

Left ventricle Clarity is a legacy attribute in objective assessment. Unlike non-medical images, apical chambers of any zoomed region can only present rough boundaries and contractive edges. Kurt, et al., (2019), have demonstrated the impact of contrast echocardiography [15], however, with respect to quantification, cavity clarity is visualized by several distinguishable fast-moving pixel's formations during cardiac cycles. This attribute, therefore, addresses the degree of distinguishable pixel element representing the endocardial border cavities or clear distinction between the intraventricular septum, valves, any trabeculated pericardial fluids and endocardial walls. Cardiac frames with very high contrast or very low contrast represent the extreme end of the spectrum and pose significant challenges [4], [20] with inexperienced operators. Equation (3) describes the root mean squared (RMS) contrast, $C_{i,j}$ which does not depend on angular frequency content or spatial distribution as best suited for 2D cardiac frames. This is given as the difference between the standard deviation of normalised pixel intensity $I_{i,j}$, and mean normalised intensity $\hat{I}$, of a given anatomical pathology; where $(i, j)$ represents the $i$-th and $j$-th element of 2D image size $M, N$; An extreme contrast could generate artefacts and potentially obscured essential anatomical details. Unfortunately, echo images with low contrast do have significant anatomical details required for clinical measurement hence the need to assess each image on the merits of clarity.

$$C_{i,j} = [\frac{1}{MN}\sum_{i=0}^{M-1}\sum_{j=0}^{N-1}(I_{i,j} - \hat{I})^2]^{\frac{1}{2}} \qquad (3)$$

*2.2.3 Depth-Gain*

Depth-gain is peculiar to 2D echocardiography, and it represents a measure of intensity of discrete signal samples of a specific region of interest. The intensity of the image signals becomes susceptible to depth changes, sector width and patient's anatomical profile. Although the use of high frequency probes can yield better resolution at shallow tissue depth penetration [21], low frequency probes give the opposite effect. Consequently, signal gain at the image apex (near field) usually possesses strong intensity of high amplitude and could become excessively low at the far field region of the cardiac frame. In the same way, excessive gain can present as pulmonary fluid in some cases [13] and images with extremely low gain attributes but bear significant anatomical details or noticeable artefact are considered in clinical practice. Equation (4) describes the intensity of reflected beam, which is associated with depth gain; where $d^2\phi$ represent the luminous flux of the infinitesimal area of source $d\Sigma$, dividing by the product of $d\Sigma$, infinitesimal solid angle $d\Omega_\xi$ and $\theta_\xi$ angle between the normal $\Omega_\xi$ to the source $d\Sigma$. While the image luminance represents the photometric measure of the pixel's intensity per unit area of light for a given area of interest. Brightness therefore is the subjective impression of the object of luminance $I_{i,j}$ and is measured in candela per square meter cd/m2. Objective model therefore assesses and score any potential introduction of artefact from excessive gain, incorrect depiction of tissues or obscurity of relevant anatomical details which is relevant for measurements.

$$I_{i,j} = \frac{d^2\phi}{d\Sigma \cdot d\Omega_\xi \cos\theta_\xi} \qquad (4)$$

*2.2.4 Foreshortening*

Apical foreshortening presents as a form of perspective deformation of the LV cavity, especially in the apex region. This deformation occurs as a result of poor image acquisition skills and could effectively alter the chamber's size and renders its volumes geometrically incongruent [22]. Apical foreshortening could occur either the systolic or diastolic cycles hence, both cycles are considered during the frame's real-time assessment. Smistad et al. (2020), have described the importance of real-time detection of apical foreshortening. For instance, foreshortening can result in inaccurate quantification of ejection fraction (EF) [14] or prevent the detection of crucial pathology, especially in the apical region. We refer to this undesirable perspective transformation $I_{(x,y,z)}$ which adds additional layer *z*, to the image's 2D plane *x,y* is expressed in terms of the homogenous transformation properties given in equation (5). In the PLAX view, however, where LV apex visibility is not required, visible apex of the LV could be taken as 'false-apex' [2], therefore counts as LV foreshortening. From a clinical standpoint, eliminating foreshortedness is paramount to optimal quantification, anatomical assessment and diagnosis of many ailments including cardiomyopathy.

$$I_{x,y,z} = \begin{bmatrix} 1 & 0 & 0 & 0 \\ 0 & 1 & 0 & 0 \\ 0 & 0 & 1 & 0 \\ 0 & 0 & -\frac{1}{d} & 1 \end{bmatrix} \begin{bmatrix} x \\ y \\ z \\ 1 \end{bmatrix} = \begin{bmatrix} x \\ y \\ z \\ -\frac{1}{d} \end{bmatrix} \Rightarrow \left( -d\frac{x}{z}, \quad d\frac{y}{z} \right) \quad (5)$$

## *2.3 Expert Annotations*

To establish the ground-truth scoring for neural network developments and testing, each echo video underwent annotation process by an experienced expert, who provided an independent score value for each quality attribute defined in Table I. The score ranges from 0 to 9 under each attribute, to allow specificity and fair assessment of A4C/PLAX apical standard. Therefore, the multi-stream architecture was trained on all four attributes simultaneously to provide normalised objective scores in the final output. Expert annotations for the echo videos were used as the quality score for all constituent frames of that video for the model developments.

## *2.4 Dataset Preparations*

The study population consisted of a random sample of 11,262 Echocardiographic studies from patients with age ranges from 17 and 85 years, who were recruited from patients who had undergone echocardiography. Three frames were randomly drawn from the video and split into training (27,028 frames), and testing (6,756 frames) sub-datasets in 80:20 ratios. Fig. 1, summarizes the frame distributions for A4C and PLAX with a categorical characteristic using experts' maximum score range values of 4.5, 6.5, 9.9 designated as poor, average, and good quality (Fig. 2) respectively. Any image with 0 score were rejected and considered unsuitable for model development.

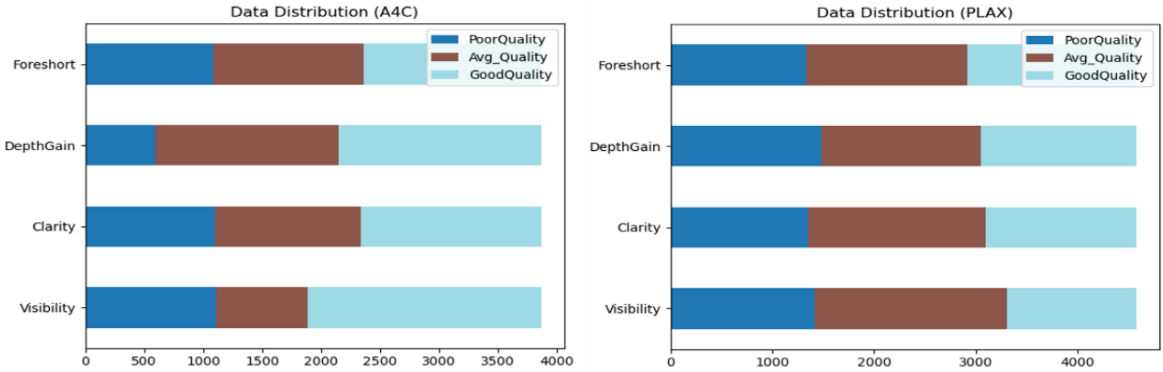

Fig. 1., Summary of data distribution for A4C and PLAX cardiac frames, indicating three category of quality levels based on experts scores values: Frames with max scores of 4.5, 6.5 and 9.9, are classified as Poor Quality, Avg Quality, and Good Quality, respectively.

**TABLE I.**
View-specific scoring definition. The quality of each view was evaluated according to several experts' elicited criteria; each criterion consisted of several attributes with independent scores but yielding a maximum score of 10 points for each criterion.

| A4C | | PLAX | |
|---|---|---|---|
| **Assessed Element per Attributes** | **Maximum Manual Score awarded** | **Assessed Element per Attributes** | **Maximum Manual Score awarded** |
| **ANATOMICAL VISIBILITY:** | | **ANATOMICAL VISIBILITY:** | |
| Correct Axis, Apical Segment | 6 | Left Ventricle (LV) Visible | **5** |
| Interventricular Septum Visible | 2 | Right Ventricle (RV) Visible | 3 |
| Interatrial Septum Visible | 2 | Full Segment Pericardium Visible | 2 |
| **ANATOMICAL CLARITY:** | | **ANATOMICAL CLARITY:** | |
| LV Cavity clarity, clear edges | 4 | LV Cavity Clarity (distinguishable border) | 4 |
| Distinguishable Valves | 3 | LV Anteroseptal Wall Clarity | 3 |
| Distinguishable Septum Wall | 3 | LV Inferolateral Wall Clarity | 3 |
| **SIGNAL DEPTH-GAIN:** | | **SIGNAL DEPTH-GAIN:** | |
| Image Sectorial Gain | **4** | Sectorial Gain | **4** |
| No Excess Gain | 3 | No Excess Gain | 3 |
| Minimum Artefacts | 3 | Minimum Artefacts | 3 |
| **LV FORESHORTEN:** | | **CAVITY FORESHORT:** | |
| LV Apical Segment present | **4** | No-Apex Diastole | 5 |
| Normal-Shaped Diastole | 3 | No-Apex Systole | 5 |
| Normal-Shaped Systole | 3 | | |

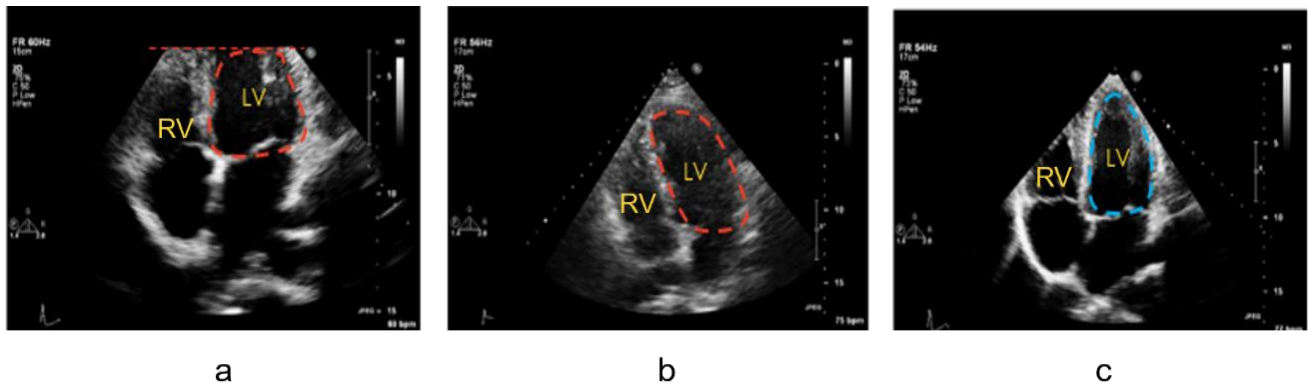

Fig 2. Showing samples of A4C with (a) cut-off apex, (b) slightly foreshortened, and (c) good quality image with clear visibility of interventricular septum, the left ventricle (LV) correctly projected (on-axis), depth-gain, and minimal foreshortening would likely gain high prediction values in quality assessment.

## 2.5 Neural Network Architecture

The architecture used in this study, referred to as 'QA-NET' is based on a multi-stream, multi-output regression model, featuring four sub-node model architectures fused together to simultaneously train and make predictions in a multi-labelled fashion. The model accepts input frame of variable length (spatial size 224×224-pixels) indicated by weight matrix $w_i^l$ and convolved with the conv layer of each parallel sub-model $F^{i-1}$ as the input feature-map to achieve

2D output feature map, $F_i^l$ of $i^{th}$ kernel of the specific conv layer $l$, given in equation (6). This flattened vector is fed into a layer of time sequence module (LSTM) for temporal extraction. Each convolution layer features an activation function of type - Rectifier Linear Units (ReLU) [23], equation (7) where neurons' activation values of $x$ ranges from 0 to maximum value. Each node was dedicated to extracting a specific anatomical feature relating to criteria defined for A4C and PLAX standard views in Table I. The components of the sub-node architecture are structurally optimized for each specific quality attribute and adapted based on best performing architecture against each quality attribute. The model features both spatial and temporal modules, illustrated in Figure 2, and detailed as follows:

(a) Each of the spatial module consists of four convolutional layers (except the clarity module with three convolutional layers), with kernel configuration of 32, 32, 32, 64 and of 3x3 size, respectively. Each convolutional layer features batch normalization [24], except the third convolutional layer which missed out on max pooling [25], and dropout of 0.5 [26]. The output is flattened, and the sequence is fed into the temporal module.
(b) Temporal module consists of an LSTM layer, used to extract temporal features. It accepts vector data from each adjacent convolutional module to compute mean score on frames' sequence data. Features are based on fast changing pixel intensity between consecutive frames, which resulted in noises where increase in varnishing gradients on training data became apparent. Therefore, output layers were configured differently to feature two stages of dense layer, batch normalization and dropouts of 0.5. This was noted to offer resilience against noisy labels and reduce variance in the image/frame data.

The choice of architecture was based on the performance data, memory requirement, and fastest inference speed data which are significant for real-time feedback implementation. We also investigated of the well-known, state-of-the-art network architectures as found in relevant resources: DenseNet121 [27], ResNet [28], VggNet [29] and compare the performance along with each 2d+t hybrid versions.

The model was trained using a 5-fold cross validation technique to ensure adequate learning on the dataset and performance was recorded for each model. The hyper parameters learning rate was set at 0.002 with high momentum 0.95 and decay rate of 0.1 every 24 steps and were reproducibly initialized to minimize possible deviation in score performance. Data augmentation was applied to allow optimum learning sequences for the models; a maximum translation of [-0.05, +0.05] pixels and maximum rotation of 5 degrees were applied randomly for horizontal, vertical, and rotational angles, respectively. To prevent overfitting in the training phase, we applied batch normalization and dropout. A multi-label optimization approach was adopted [30], and the model was trained simultaneously using four quality attributes with mean absolute error as the cost function. Training was initialized with 32 batch size and completed as learning curves converged around 40 epochs.

$$F_{i,jk}^l = \sum_{i=0}^{n}\sum_{j=0}^{m} w_{i,mn}^l \, F_{(j+m)(k+n)}^{l-1} \qquad (6)$$

$$f(x)_{relu} = max(0,x) \qquad (7)$$

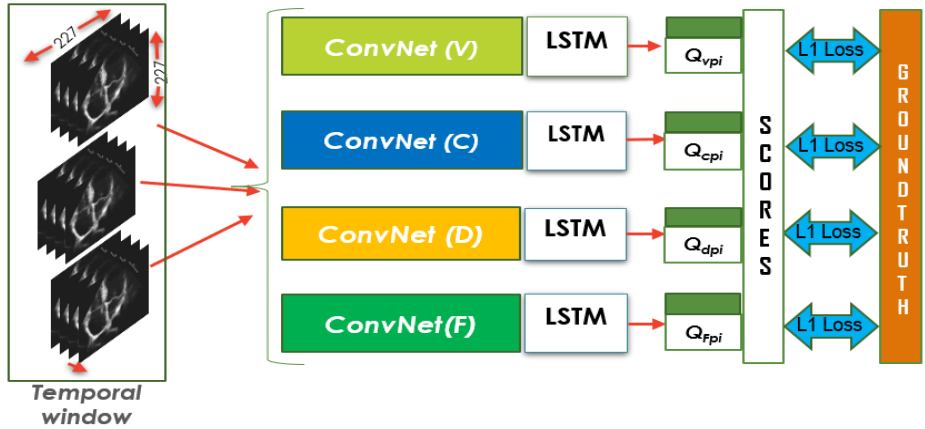

Figure 2. The multi-stream neural network architecture referred to as 'Q-NET'. Each stream is dedicated to specific prediction and assessment of images quality on visibility, clarity, depth-gain and foreshortedness as identified by $Q_V$, $Q_C$, $Q_D$, and $Q_F$, respectively.

## 2.6 Evaluation Metrics

Since the model uses multiplex variables for each score attributes, the output score was normalised to [0 ~ 1] via sigmoid activation function $f(x)$, equation (8) and prediction error were evaluated against the MAE, taking the average of the absolute difference between cardiologist's ground truth ($Q_{GT}$) scores and model's predicted scores ($Q_P$). Therefore, model's minimal error with values close to 0, would indicate a best fit scenario while larger error with values close to 1, would indicate a poor fit regression model. Lastly, the average model's performance in percentage was computed in equation (9).

$$f(x)_{sigmoid} = 1/(1 + e^{-x}) \quad (8)$$

$$Model_{acc} = 1 - \left(\frac{\sum_{i=0}^{n}|Q_{GTi} - Q_{pi}|}{n}\right) * 100 \quad (9)$$

## 3. RESULTS

Cardiac echo frames are laced with significant complexities among which are patients anatomical and pathological differences, these complexities are reflected in each fast-moving echo frame therefore, the model's inference speed is overly critical to the real-time assessment and operators' feedback guidance for possible optimization. Hence, implementing a customized model that can successfully generalize with high confidence and high-speed inference would make a significant achievement in automated assessment.

The proposed multi-stream model was evaluated on external dataset to lower systemic bias and achieved a mean accuracy of 96.20% and 2.52 ms inference speed which reinforces the viability for real-time feedback deployment per quality per frame. The results enumerated in Table II outlined the accuracies and error distribution achieved on each dedicated model for visibility, clarity, depth-gain and foreshorten indicators respectively. The aggregated error distribution per model is depicted in Figure 3, while the model predictive accuracy (Table III) for visibility, clarity, depth-gain, and foreshortening group attributes (indicators) are 94.4%, 96.8%, 96.2% and 97.4% respectively. The samples of predicted cardiac frames shown in Fig. 4, clearly indicates the predictive objective scores for visibility, clarity, depth-gain, foreshortedness and its weighted average score (AS). These are automatically generated by the pipeline and superimposed on the cardiac frames in real-time.

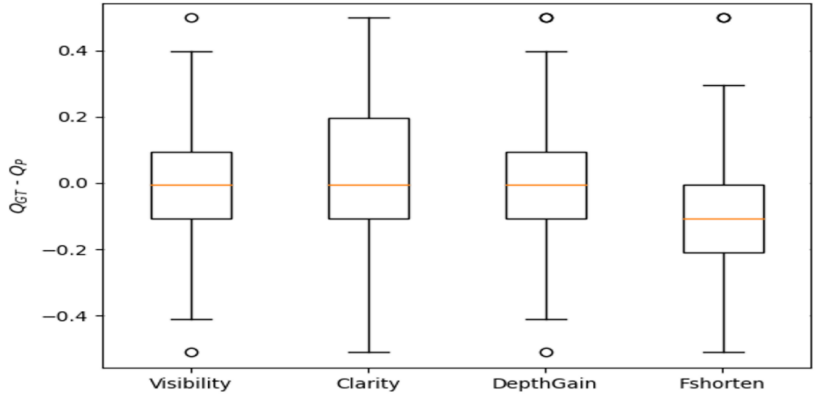

Fig. 3. Box plot showing the error distribution on the multi-stream model architecture featuring the collective quality attributes indicators. The y-axis showing the difference between experts ground truth and model predictions. The x-axis showing each quality attributes (indicator) per model.

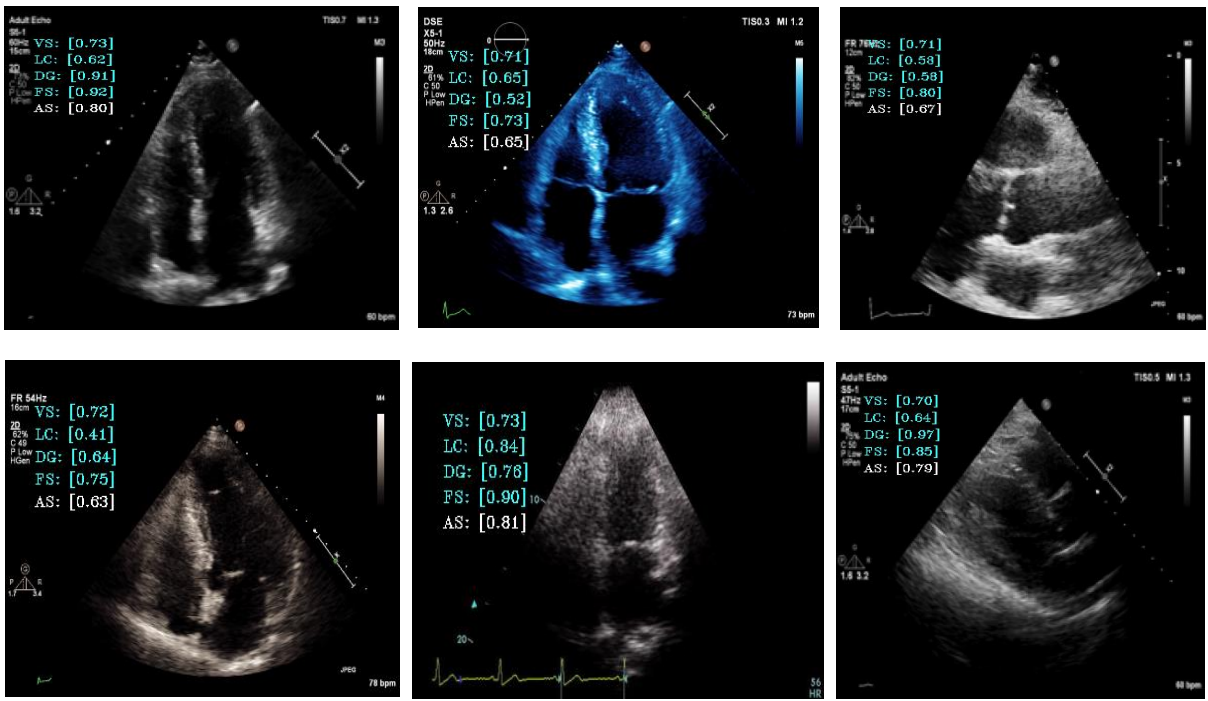

Fig. 4. Sample of predicted images with respective objective scores. visibility (VS), clarity (LC), depth gain (DG), foreshortening (FS) and overall quality score (AS) are used to assess cardiac image quality during clinical acquisition process.

TABLE II
Model performance on quality attribute/indicator and the error distribution, given by the mean and deviation (μ ± σ) notation. Q1, Q2, and Q3 represent the three levels of image quality per model. Inference time is the average time it takes the model to identify and predict scores for each image's quality levels.

| Quality Model / Indicator | Acc.% | Q1 | Q2 | Q3 | Inference Time (ms) |
|---|---|---|---|---|---|
| | | μ ± σ | | | |
| **Visibility** | 94.37 | 0.1536 ± 0.1036 | 0.1681 ± 0.1100 | 0.1581 ± 0.1038 | 9.476 |
| **Clarity** | 96.84 | 0.1635 ± 0.1034 | 0.1759 ± 0.1042 | 0.1837 ± 0.1106 | 7.753 |
| **Depth-Gain** | 96.27 | 0.1357 ± 0.0982 | 0.1326 ± 0.0956 | 0.1306 ± 0.0956 | 8.384 |
| **Foreshorten** | 97.50 | 0.1909 ± 0.1262 | 0.1901 ± 0.1285 | 0.2044 ± 0.1242 | 9.356 |

TABLE III
Comparison of the model's performance, including the selected state-of-the-art model (DenseNet, ResNet and VggNet) on echocardiogram objective quality score. Each of the model were evaluated on the combined attributes (quality indicator) of visibility, clarity, depth-gain, and foreshortening simultaneously. The best performing model (QA-Net), in terms of accuracy and inference is highlighted.

| MODEL DATA | REGRESSION ACCURACY (%) | | | | | Inference Time (ms) |
|---|---|---|---|---|---|---|
| | Visibility | Clarity | Depth-Gain | Foreshorten | Accuracy | |
| DENSNET+LSTM | 92.20 | 88.2 | 95.62 | 90.44 | 91.62 | 24.704 |
| RESNET+LSTM | 89.45 | 92.25 | 87.40 | 92.20 | 90.32 | 19.526 |
| VGGNET+LSTM | 92.30 | 97.20 | 98.40 | 89.20 | 94.28 | 30.760 |
| **QA-NET+LSTM** | **94.40** | **96.80** | **96.20** | **97.40** | **96.20** | **2.52** |

## 4. DISCUSSION

The results of each model's performance (except QA-NET+LSTM) shown in Table III, vary substantially on group attribute/indicator even though each model retains its original values of hyperparameters. This indicatively prove that one model cannot fit it all. Each of the evaluated state-of-the-art model is incapable of delivering consistent performance in terms of accuracy, real-time inference speed on the selected group quality attributes/indicators.

The achieved inference speeds for DenseNet121, ResNet50, and VggNet16, are 24.70ms, 19.53ms, and 30.76ms, respectively. This implies that a maximum frame per second (FPS) of 40, 51 and 32 can be achieved with the respective state of the art models. This speed is insufficient for real-time quality assessment. Nevertheless, our model (QA-Net) achieved a reduction of 90% in inference speed for the combined model performance, making it the best candidate for transthoracic image quality assessment solution. Note that our multi-stream pipeline actively combined four attributes group indicators simultaneously during the training and prediction phases. The deep neural networks were fully optimised to yield the best inference speed and performance accuracy on each of the specific quality indicator. Therefore, we concluded that each quality attribute/indicator associated with distinguishable pathological complexities and fast-changing echocardiogram would require a fully customized and optimized model for objective quality assessment task.

Furthermore, a summary of visualization of the learned features is illustrated in Fig 5. To obtain this, each quality indicator group was modelled individually to show the discriminative ability of the respective network of our model. Here the model's feature maps give the idea on which part of image's element is being focused upon at each convolutional layer in the network. Although, several global characteristics have been used in the quality criteria, yet this study does not claim exhaustiveness in the group criteria indicators. We are aware that different laboratories are at liberty to adopt what is considered the best practice in their region of practice especially when such requirements are mandatory by the healthcare legislation.

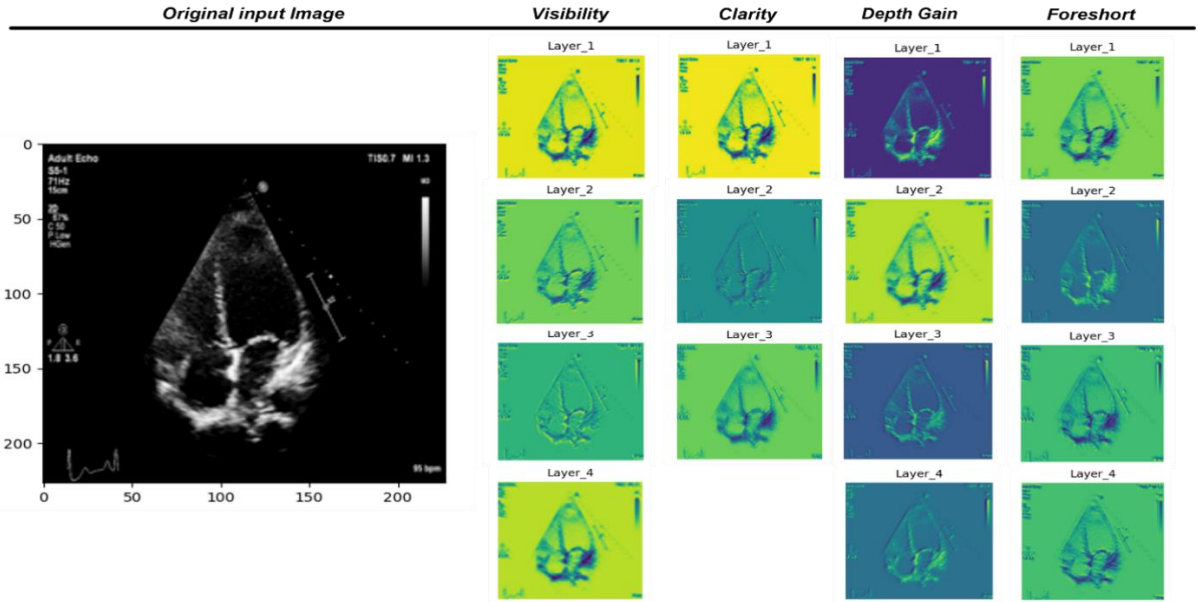

Fig 5. Illustrating the convolutional layers learned features (feature map) from a specific image (original input image) and its respective layers of distinguishable detection from each model.

## 5. CONCLUSION

In this study, we have considered four distinctive group of quality attributes or indicators where echocardiographic image quality was evaluated and a novel method of accessing such attributes under A4C and PLAX echocardiographic views. These two views bear clinical significance and are recommended for chamber quantification and linear measurement in clinical workflow [1], [2], [3]. Since this appear to be the first time where 2D echocardiogram quality indicators are thus defined comprehensively, the results of our work can only be compared to the recent, and existing work on echocardiographic image quality assessment listed in Table IV. Technical comparison, in terms of the use case functionality, clinical feasibility, and method of assessment could provide significant evidence on how this important clinical problem in echocardiographic image quality has been addressed. Furthermore, each of the existing work (including ours) have employed different independent dataset, study population and sample size. Even though our works achieved better model accuracy compared to all the existing similar works, it is reasonably unfair to compare the model accuracy or inference speed event, except on the clinical deployability and use case functionality.

We have presented the clinical significance and feasibility of developing an automated quality assessment in two-dimensional echocardiographic images. Therefore, a quantitative method defined for image quality standard can provide useful feedback for an operator guidance system as well as a valuable tool for research in clinical practice, first to function as arbiter reference to clinicians, secondly, to accelerate the learning curve for those in training. Also, it can provide specific information on the adequacy of the images obtained in retrospect, which could be universally relevant for a lifesaving procedure at point of care or during clinical emergencies.

Finally, we used the annotation provided by two experts, a cardiologist who provided reference and supervision and an accredited annotator. An intra-observer variability can be examined by obtaining additional annotations from human experts and compared with the error in the predicted scores.

TABLE IV
Summary of model performance on quality assessment pipeline in related studies

| **Studies by Author** | **Abdi et al., (2017)** | **Luong et al., (2020)** | **Dong et al., (2020)** | **Current Study (2022)** |
|---|---|---|---|---|
| **Ultrasound Source** | Philips and GE | Philips iE33 platform Philips S51 frequency 5–1 MHz | Shenzhen Maternal and Child Healthcare Hospital | GE Healthcare (Vivid.i) and Philips Healthcare (iE33 xMATRIX) |
| **Study Population** | N/A | 3,157 Patients | N/A | 11,262 Patients |
| **Ground Truth annotations** | 2 expert annotations | 1 level 3 echo cardiographer | 1 radiological Expert | 4 annotations each by 2 AV experts |
| **Quality Criteria** | 13 | 12 | 6 | **23** (Table I) |
| **# Of image quality attributes and Assessment Methods** | 1 (WAS) Method grossly insufficient of clinical advantage | 1 (WAS) Method grossly insufficient of clinical advantage | 2 (Zoom, Gain) Focus on fetal ultrasound, adult not implemented | **4** (Visibility, Clarity, Depth Gain & Foreshortening) Relevant to TTEs |
| **# of standard views considered and type** | 5 (A2C, A3C, A4C, $PSAX_A$, $PSAX_{PM}$) | 9 (PLAX, A2C, A3C, A4C, PSAX-A, PSAX-M, PSAX-PM, SC4 & IVC) | 1 (A4C only) | 2 A4C & PLAX *(Apical views relevant to TTE standards)* |
| **Input Size** | 200 x 200 x 3 | Not specified | 224 x 224 x 3 | 224 x 224 x 3 |
| **Sample Size** | 6,916 | 14,086 | 2,032 | 33,784 |
| **Model Accuracy achieved** | **85.0%** | **87.0%** | **93.52%** | **96.2%** |

**Funding:** This research did not receive any specific grant from funding agencies in the public, commercial, or not-for-profit sectors.

**Authors Contributions:** Robert Labs: original draft preparation, data curation, writing; Apostolos Vrettos: ground truth annotations; Jonathan Loo: validation, reviewing; Massoud Zolgharni: reviewing, formatting, editing.

**Conflict of Interest:** The authors declare that there are no conflicts of interest.